\documentclass[12pt]{article}
\pdfoutput=1
\usepackage{amssymb}
\usepackage{amsmath}
\usepackage{latexsym}
\usepackage[usenames]{color}
\usepackage[mathscr]{eucal}
\usepackage{fancybox}
\usepackage{simplewick}
\usepackage{comment}
\usepackage{cite}
\usepackage{bm}
\usepackage{framed}
\definecolor{shadecolor}{rgb}{0.9,0.9,0.95}
\usepackage{setspace}
\usepackage[normalem]{ulem}
\definecolor{darkgreen}{rgb}{0,0.5,0}
\definecolor{darkblue}{cmyk}{0.9,0.9,0,0}
\definecolor{darkred}{rgb}{0.6,0,0.3}

\usepackage{graphicx}
\usepackage[curve,matrix,arrow,color]{xy}
\usepackage{tikz}
\usetikzlibrary{intersections}
\usetikzlibrary{er,positioning}
\usetikzlibrary{arrows,shapes}
\usetikzlibrary{decorations.markings}
\usepackage[setpagesize=false,pagebackref=false, linktocpage, bookmarksopen=true, colorlinks=true, linkcolor=darkblue,citecolor=darkblue,urlcolor=black]{hyperref}
\usepackage{hyperref}

\renewcommand{\thefootnote}{\arabic{footnote}}

\def\eqref#1{(\ref{#1})}

\def\beq{\begin{equation}}
\def\eeq{\end{equation}}

\numberwithin{equation}{section}

\begin{document}
\thispagestyle{empty}

\renewcommand{\thefootnote}{\fnsymbol{footnote}}
\setcounter{page}{1}
\setcounter{footnote}{0}
\setcounter{figure}{0}
\vspace{0.7cm}
\begin{center}
\Large{$T\bar{T}$-deformed Entanglement Entropy for IQFT}
\vspace{1.3cm}

\normalsize{Miao He, Jue Hou, Yunfeng Jiang}
\\ \vspace{1cm}
\footnotesize{\textit{
School of physics \& Shing-Tung Yau Center, Southeast University,\\ Nanjing  211189, P. R. China
}
}
\par\vspace{1.0cm}
\textbf{Abstract}\vspace{2mm}
\end{center}
\noindent
We calculate the $T\bar{T}$-deformed entanglement entropy for integrable quantum field theories (IQFTs) using the form factor bootstrap approach. We solve the form factor bootstrap axioms for the branch-point twist fields and obtain the deformed form factors. Using these form factors, we compute the deformed von Neuman entropy up to two particle contributions. The solution of the form factor axioms is not unique. We find that for the simplest solution of the bootstrap axioms, the UV limit of the entanglement entropy takes the same form as the undeformed one, but the effective central charge is deformed. For solutions with additional CDD-like factors, we can have different behaviors. The IR corrections, which only depends on the particle spectrum is untouched. 
\setcounter{page}{1}
\renewcommand{\thefootnote}{\arabic{footnote}}
\setcounter{footnote}{0}
\setcounter{tocdepth}{2}
\newpage
\tableofcontents

\section{Introduction}
\label{sec:intro}
$T\bar{T}$-deformation and other related irrelevant solvable deformations \cite{Smirnov:2016lqw,Cavaglia:2016oda} have received considerable interest in recent years. Such deformations are under much better analytic control than a generic irrelevant deformation. The deformed theories are no longer usual local QFTs that we learn in textbooks, but exhibit intriguing new UV behaviors that resembles gravity theory \cite{Dubovsky:2017cnj,Dubovsky:2018bmo,Cardy:2018sdv,Coleman:2019dvf,Tolley:2019nmm} and string theory \cite{Callebaut:2019omt,Benjamin:2023nts,Baggio:2018gct,Sfondrini:2019smd}. The interests on these deformations ranges from holography \cite{McGough:2016lol,Guica:2019nzm,Kraus:2018xrn,Hartman:2018tkw,Taylor:2018xcy,Caputa:2019pam,Caputa:2020lpa,Gorbenko:2018oov,Lewkowycz:2019xse,Gross:2019ach,Gross:2019uxi,Giveon:2017nie,Chakraborty:2020swe,Chakraborty:2020cgo,Giveon:2017myj,Apolo:2018qpq,Apolo:2019zai,Apolo:2021wcn,Apolo:2023vnm,Li:2020pwa,Hirano:2020nwq} to integrable models \cite{Pozsgay:2019ekd,Marchetto:2019yyt,Aramini:2022wbn,Ceschin:2020jto,Conti:2019dxg,Conti:2018tca,Jiang:2020nnb,Hansen:2020hrs,Doyon:2021tzy,Cardy:2020olv,Medenjak:2020bpe,Medenjak:2020ppv,Ahn:2022pia,LeClair:2021wfd}.\par

Solvability of such deformations imply that many physical quantities can be computed exactly. However, there is a sharp distinction between global quantities such as the spectrum \cite{Smirnov:2016lqw,Cavaglia:2016oda}, the S-matrix \cite{Smirnov:2016lqw,Dubovsky:2017cnj}, the torus partition functions \cite{Cardy:2018sdv,Dubovsky:2018bmo,Datta:2018thy,Aharony:2018bad,Aharony:2018ics} and the quantities involving local operators such as multi-point correlation functions. The former are relatively easier to compute and indeed much progress have been made in the early stage of the study of $T\bar{T}$-deformation. Correlation functions are more challenging due to various ambiguities and lack of non-perturbative techniques. Some features can be obtained in certain special limits or by performing perturbative calculations up to first few orders \cite{Kraus:2018xrn,Aharony:2018vux,He:2019ahx,He:2020udl,He:2020qcs,Hirano:2020ppu,He:2023hoj}. An early non-perturbative investigation of the problem was performed by Cardy using random geometry \cite{Cardy:2019qao}. A non-perturbative approach based on deformed symmetry has been proposed by Guica in \cite{Guica:2021fkv}.\par
Very recently, the situation for computing deformed correlation functions non-perturbatively has improved drastically thanks to several developments. In \cite{Cui:2023jrb}, the authors computed deformed two-point functions for the single-trace $T\bar{T}$-deformed orbifold CFT. The computation is done by holography using worldsheet techniques and checked on the field theory side. In the un-twisted sector, it is believed that the result should be the same as the usual double-trace $T\bar{T}$-deformation. This is indeed confirmed by \cite{Aharony:2023dod}  where the authors take a different approach by using JT gravity formulation of the $T\bar{T}$ deformation. The results from these two different approaches are indeed consistent with each other.\par
These works mainly concern the $T\bar{T}$-deformed CFTs. Another class of theories which are under good analytic control are the massive integrable QFTs. For these theories, a standard approach to compute correlation functions is the form factor bootstrap approach \cite{Karowski:1978vz,Smirnov:1992vz,Delfino:2003yr}. It has been suggested that the deformed correlation functions can be studied in this way in the original work of Smirnov and Zamolodchikov \cite{Smirnov:2016lqw}. This was recently carried out systematically in the nice works \cite{Castro-Alvaredo:2023rtl,Castro-Alvaredo:2023wmw,Castro-Alvaredo:2023hap}. The authors solved the deformed form factor bootstrap axioms for form factors of local and quai-local operators for IQFTs with diagonal S-matrices. It turns out that the deformed form factors take factorized forms, similar to the S-matrix. Based on the results, they checked the deformed $\Delta$-sum rule and analyzed in some detail the UV behavior of the two-point functions.\par
Building upon these works, we shall take one step further and compute the quantum entanglement entropy for IQFTs. Previous works on the computation of entanglement of $T\bar{T}$-deformed CFT form both field theory and holography can be found in \cite{Donnelly:2018bef,Banerjee:2019ewu,Grieninger:2019zts,Donnelly:2019pie,Lewkowycz:2019xse,Chen:2018eqk,Sun:2019ijq,Jeong:2019ylz,He:2019vzf,Chakraborty:2018kpr,Chakraborty:2020udr,Asrat:2020uib,Jeong:2022jmp,Allameh:2021moy,He:2023xnb,Tian:2023fgf}. The main idea is to compute the two-point function of branch-point twist fields which takes into account the structure of the $n$-sheet Riemann surface which arise in the replica trick. Since the branch-point twist fields are non-local, their form factor axioms are modified. The form factor bootstrap program for these fields have been proposed in \cite{Cardy:2007mb}. We will solve the bootstrap program for the more general CDD factor-deformed theories for the branch-point twist field. The deformed form factors are then used to compute the  deformed scaling dimension of the twist operators, assuming the validity of $\Delta$-sum rule in the deformed theories. The two-point function of the twist operators gives the entanglement entropy. The deformed scaling dimension of the twist operator in the $n\to 1$ limit plays a key role in the UV limit of the entanglement entropy. We find that although the scaling dimension of the twist field is indeed deformed for generic $n$, in the $n\to 1$ limit the scaling dimension still vanishes, as least for the simplest solution of the form factor axioms. An immediate consequence of this fact is that the UV limit of entanglement entropy takes the same form as the undeformed one, but with a deformed effective central charge.\par
Before going into technical details, let us make an important remark about the solution of the form factor axioms. The bootstrap axioms is a set of postulates that one believes the form factors should satisfy based on general principles. For local IQFTs, such postulates can be well justified based on general properties of local QFTs \cite{Smirnov:1992vz,Babujian:1998uw} and has passed many non-trivial tests. The solutions of the bootstrap axioms are not unique. To fix the form factor of a given operator, one needs extra input such as vacuum expectation value of the operator and asymptotic behavior of the form factors. In addition, one can make cross check of the correlation functions with other non-perturbative approaches such as the truncated conformal space approach \cite{Feverati:1998va,Yurov:1989yu,Yurov:1990kv}.
\par
For $T\bar{T}$-deformed theories, the situation is different. Since these theories are not usual local QFTs, in principle one needs to justify the validity of the bootstrap axioms. Although we believe this is a reasonable assumption, given that other integrability-based approaches like thermodynamic Bethe ansatz work well for the spectral problem. Another difficulty is that at the current stage, we do not have enough data to fix the ambiguity of the solution of the bootstrap axioms. The asymptotic behavior of the form factors is based on the knowledge of the asymptotic behaviors of two-point functions, which is not completely understood in the $T\bar{T}$-deformed theories yet. The vacuum expectation value of local operators in the deformed theory poses another difficulty. Therefore, the solutions of the form factors for the branch-point twist fields found in this paper are the simplest solutions, which might not be the `correct solutions'. Nevertheless, we believe the results in this paper can still be useful as it constitute the first step towards the correct answer.
\par
The rest of the paper is organized as follows. In section~\ref{sec:FF}, we review the form factor bootstrap program for the branch-point twist fields and solve the deformed form factors. In section~\ref{sec:UVdim}, we compute the deformed scaling dimension for the branch-point twist field. In section~\ref{sec:EE}, we compute the deformed entanglement entropy and analyze the results. We conclude in section~\ref{sec:conlusion} and discuss future directions.

\section{Form factor bootstrap for twist operators}
\label{sec:FF}
Let us consider an integrable QFT in 1+1 dimensions. We consider the ground state $|0\rangle$ in the infinite volume and take $A$ to be a finite interval. The full Hilbert space is $\mathcal{H}=\mathcal{H}_A\otimes\mathcal{H}_B$ where $B$ is the complement of $A$. The bipartite entanglement entropy is defined as
\begin{align}
\rho_A=\text{Tr}_{\mathcal{H}_B}\left( |0\rangle\langle0|\right),\qquad
S_A=-\text{Tr}_{\mathcal{H}_A}\left( \rho_A\log(\rho_A)\right)\,.
\end{align}
\par

We compute the entanglement entropy by the replica trick. In this method, we make $n$ copies of the theory. We first compute the partition function on a $n$-sheet Riemann surface where the branch points correspond to the end points of $A$. After computing the partition function, one then perform an analytic continuation in $n$ and finally take the limit $n\to 1$. The partition function of the $n$-sheet Riemann surface can be computed by the two-point function of the so-called branch-point twist operators $\mathcal{T}$, $\bar{\mathcal{T}}$ located on branch points, which boosts the fundamental fields from one sheet to another. The two-point function can be computed by the form factor approach. But due to the special properties of the branch-point twist field, the form factor bootstrap axioms have to be modified. In this section, we first review the form factor axioms for the branch-point twist fields and then solve the deformed form factors under a general CDD factor deformation \cite{Castro-Alvaredo:2023rtl}. We will focus on the $T\bar{T}$ deformation when we compute entanglement entropy.
\subsection{Bootstrap axioms}
\label{sec:3.1}
We consider $n$ copies of a known integrable theory possessing a single particle spectrum and no bound states such as the Ising and sinh-Gordon models. We have therefore $n$ kinds of particles, which we will denote by indices $1,2,...,n$. The S-matrix between particles $i$ and $j$ with rapidities $\theta_i$ and $\theta_j$ is denoted by $S_{ij}(\theta_i-\theta_j)$.  We introduce the Fadeev-Zamolodchikov operators on each sheet and denote them by $Z_{i}^{\dagger}(\theta_{i})$ and $Z_{i}(\theta_{i})$. The operators on different copies are commutative. We can package the $n$-copy theory into one theory, which contains $n$ kinds of particles. They  satisfy the Fadeev-Zamolodchikov algebra
\begin{align}
Z_{i}^{\dagger}(\theta_i)Z^{\dagger}_{j}(\theta_{j})&=S_{ij}(\theta_i-\theta_j)Z_j^\dagger(\theta_j)Z_i^\dagger(\theta_i), \\
Z_{i}(\theta_i)Z_{j}(\theta_{j})&=S_{ij}(\theta_i-\theta_j)Z_j(\theta_j)Z_i(\theta_i), \\
Z_i(\theta_i)Z_j^{\dagger}(\theta_j)& =S_{ij}(\theta_i-\theta_j)Z_j^\dagger(\theta_j)Z_i(\theta_i)+\delta(\theta_i-\theta_j),
\end{align}
where the $S$-matrix is defined by
\begin{align}
&S_{ij}(\theta)=\left\{\begin{array}{lll}
S(\theta)&\quad i=j\quad i=1,\ldots,n\\ 1&\quad i\neq j\quad i,j=1,\ldots,n\end{array}\right.
\end{align}
Here $S(\theta)$ is the S-matrix of the single-copy integrable QFT.
\par
The form factor for the theory on $n$-sheet Riemann surface is defined by
\begin{align}
F_k^{\mathcal{O}|\mu_1...\mu_k}(\theta_1,\ldots,\theta_k):=\langle0|\mathcal{O}|\theta_1,\ldots,\theta_k\rangle_{\mu_1\ldots\mu_k}
\end{align}
where the $k$-particle asymptotic state is given by
\begin{align}
|\theta_1,\ldots,\theta_k\rangle_{\mu_1,\ldots\mu_k}=Z^{\dagger}_{\mu_1}(\theta_1)...Z^{\dagger}_{\mu_k}(\theta_k)|0\rangle.
\end{align}
The index $\mu_i$ means the $i$-th particle is created by $Z^{\dagger}_{\mu_i}$. $(\mu_i\in \{1,2,...,n\})$
\par
Now we list the form factor axioms for the branch-point twist fields~\cite{Cardy:2007mb}
\begin{itemize}
\item Watson's equation
\begin{align}
F_k^{\mathcal{T}|{\dots\mu_i\mu_{i+1}\dots}}(\dots,\theta_i,\theta_{i+1},\dots)=&S_{\mu_i\mu_{i+1}}(\theta_{i}-\theta_{i+1})F_k^{\mathcal{T}|{\dots\mu_{i+1}\mu_i\dots}}(\dots,\theta_{i+1},\theta_i,\dots);
\end{align}
\item Cyclicity
\begin{align}
F_k^{\mathcal{T}|\mu_1\mu_2...\mu_k}(\theta_1+2\pi i,\ldots,\theta_k)=F_k^{\mathcal{T}|\mu_2...\mu_n\hat{\mu}_1}(\theta_2,\ldots,\theta_k,\theta_1);
\end{align}
where $\hat{\mu}_i=\mu_i+1$\,.
\item Kinematic poles
\begin{align}
\label{eq:krPole}
\lim_{\bar{\theta}\to\theta}(\bar{\theta}\to\theta)F_{k+2}^{\mathcal{T}|\bar{\mu}\mu\mu_1\ldots\mu_k}(\bar{\theta}+i\pi,\theta,\theta_1\ldots,\theta_k)&=iF_k^{T|\mu_1\ldots\mu_k}(\theta_1,\ldots,\theta_k),  \\\nonumber
\lim_{\bar{\theta}\to\theta}(\bar{\theta}\to\theta)F_{k+2}^{T|\bar{\mu}\hat{\mu}\mu_{1}\ldots\mu_{k}}(\bar{\theta}+i\pi,\theta,\theta_{1}\ldots,\theta_{k})&=-i\prod_{i=1}^{k}S_{\mu\mu_{i}}(\theta-\theta_{i})F_{k}^{T|\mu_{1}\ldots\mu_{k}}(\theta_{1},\ldots,\theta_{k}).
\end{align}
\end{itemize}
In principle we also have axioms for the bound states. Since we are focusing on theories without bound states, they are not relevant here.
\paragraph{Minimal form factor} To solve these equations, we first construct the so-called minimal form factor, satisfying
\begin{align}
F_{\min}^{\mathcal{T}|kj}(\theta,n)=F_{\min}^{\mathcal{T}|jk}(-\theta,n)S_{kj}(\theta)=F_{\min}^{\mathcal{T}|jk+1}(2\pi i-\theta,n)\quad\forall\quad j,k
\end{align}
Repeated use of the above equations leads to the following constraints:
\begin{align}
F_{\min}^{\mathcal{T}|ii+k}(\theta,n)&=F_{\min}^{\mathcal{T}|jj+k}(\theta,n)\quad\forall\quad i,j,k  \\
F_{\min}^{\mathcal{T}|1j}(\theta,n)& =F_{\min}^{\mathcal{T}|11}(2\pi(j-1)i-\theta,n)\quad\forall\quad j\neq1. 
\end{align}
Therefore, the form factor $F_{\min}^{T|11}(\theta,n)$ is enough to determine all minimal form factors of the theory, which also satisfies
\begin{align}
F_{\min}^{\mathcal{T}|11}(\theta,n)=F_{\min}^{\mathcal{T}|11}(-\theta,n)S(\theta)=F_{\min}^{\mathcal{T}|11}(-\theta+2\pi ni,n).
\end{align}
Recall that the minimal form factor for local operators satisfy
\begin{align}
\label{eq:undeformed F relation}
F_{\min}(\theta)=F_{\min}(-\theta)S(n\theta)=F_{\min}(-\theta+2\pi i),
\end{align} 
which allows us to identify the solution
\begin{align}
F_{\min}^{\mathcal{T}|11}(\theta,n)=F_{\min}(\theta/n).
\end{align}
\paragraph{The two-particle form factor} To determine the full two-particle form factor, we also need to include poles in the extended physical strip. The pole structure is determined by the kinematic residue equations \eqref{eq:krPole}. It turns out that the full form factor has two poles in the extended physical sheet at $\theta=i\pi$ and $\theta=i\pi(2n-1)$, which leads to
\begin{align}
F_2^{\mathcal{T}|jk}(\theta,n)=\frac{\langle \mathcal{T}\rangle\sin\left(\frac{\pi}{n}\right)}{2n\sinh\left(\frac{i\pi(2(j-k)-1)+\theta}{2n}\right)\sinh\left(\frac{i\pi(2(k-j)-1)-\theta}{2n}\right)}\frac{F_{\min}^{\mathcal{T}|jk}(\theta,n)}{F_{\min}^{\mathcal{T}|jk}(i\pi,n)},
\end{align}
where the normalization has been chosen so that the kinematical residue equation gives
\begin{align}
F_0^{\mathcal{T}}=\langle\mathcal{T}\rangle.
\end{align}
\paragraph{Twist operator $\mathcal{\tilde{T}}$}
For the branch-point twist field $\mathcal{\tilde{T}}$, Watson's equation implies that form factors of the operator $\mathcal{\tilde{T}}$ are equal to the form factor of $\mathcal{T}$ up to the transformation $i\to n-i$ for each particle
\begin{align}
F_2^{\mathcal{T}|ij}(\theta,n)=F_2^{\mathcal{\tilde{T}}|(n-i)(n-j)}(\theta,n).
\end{align}
This leads to
\begin{align}
F_2^{\mathcal{\tilde{T}}|11}(\theta,n)&=F_2^{\mathcal{T}|11}(\theta,n),  \\
F_{2}^{\mathcal{\tilde{T}}|1j}(\theta,n)&=F_2^{\mathcal{T}|11}(\theta+2\pi i(j-1),n). 
\end{align}
Higher particle form factors can be determined following the lines \cite{Castro-Alvaredo:2011hnn}. As a first step, we consider up to two-particle contributions in this paper and will leave higher particle contributions for future study.
\subsection{Turning on deformations}
\label{sec:3.2}
Now we consider the CDD factor deformation, which means we deform the S-matrix of the IQFT by multiplying a CDD factor
\begin{align}
S_{\boldsymbol{\alpha}}(\theta)=\Phi_{\boldsymbol{\alpha}}(\theta)S(\theta),\quad \Phi_{\boldsymbol{\alpha}}(\theta)=\exp\left[-i\sum\limits_{s\in\mathcal{S}}\alpha_s\sinh(s\theta)\right].
\end{align}
The S-matrix for the $n$-copy theory becomes
\begin{align}
&S^{\boldsymbol{\alpha}}_{ij}(\theta)=\left\{\begin{array}{lll}S_{\boldsymbol{\alpha}}(\theta)&\quad i=j\quad i=1,\ldots,n\\ 1&\quad i\neq j\quad i,j=1,\ldots,n\end{array}\right.
\end{align}
For the CDD factor deformed form factor, like for the local operators, we assume that the axioms for the branch-point twist fields still hold. Notice that, for the $n$-copy theory, the full S-matrix is not simply modified by multiplying a global CDD factor, because even in the deformed theory, particles from different sheets do not interact. Therefore $S_{ij}^{\boldsymbol{\alpha}}(\theta)=1$ is not modified for $i\ne j$. This is different from the local operators. The deformed form factor axioms for branch-point twist fields read:
\begin{itemize}
\item Watson's equation
\begin{align}
F_k^{\mathcal{T}|{\dots\mu_i\mu_{i+1}\dots}}(\dots,\theta_i,\theta_{i+1},\dots;\boldsymbol{\alpha})=&S^{\boldsymbol{\alpha}}_{\mu_i\mu_{i+1}}(\theta_{i}-\theta_{i+1})F_k^{\mathcal{T}|{\dots\mu_{i+1}\mu_i\dots}}(\dots,\theta_{i+1},\theta_i,\dots;\boldsymbol{\alpha});
\end{align}
\item Cyclicity 
\begin{align}
F_k^{\mathcal{T}|\mu_1\mu_2...\mu_k}(\theta_1+2\pi i,\ldots,\theta_k;\boldsymbol{\alpha})=F_k^{\mathcal{T}|\mu_2...\mu_n\hat{\mu}_1}(\theta_2,\ldots,\theta_k,\theta_1;\boldsymbol{\alpha});
\end{align}
\item Kinematic poles
\begin{align}
\lim_{\bar{\theta}\to\theta}(\bar{\theta}\to\theta)F_{k+2}^{\mathcal{T}|\bar{\mu}\mu\mu_1\ldots\mu_k}(\bar{\theta}+i\pi,\theta,\theta_1\ldots,\theta_k;\boldsymbol{\alpha})&=iF_k^{T|\mu_1\ldots\mu_k}(\theta_1,\ldots,\theta_k;\boldsymbol{\alpha}),  \\\nonumber
\lim_{\bar{\theta}\to\theta}(\bar{\theta}\to\theta)F_{k+2}^{T|\bar{\mu}\hat{\mu}\mu_{1}\ldots\mu_{k}}(\bar{\theta}+i\pi,\theta,\theta_{1}\ldots,\theta_{k};\boldsymbol{\alpha})&=-i\prod_{i=1}^{k}S^{\boldsymbol{\alpha}}_{\mu\mu_{i}}(\theta-\theta_{i})F_{k}^{T|\mu_{1}\ldots\mu_{k}}(\theta_{1},\ldots,\theta_{k};\boldsymbol{\alpha}).
\end{align}
\end{itemize}

We can follow the undeformed case to solve the deformed two-particle form factor.
\paragraph{Minimal form factor} The deformed minimal form factor should satisfy
\begin{align}
F_{\min}^{\mathcal{T}|jk}(\theta,n;\boldsymbol{\alpha})=F_{\min}^{\mathcal{T}|jk}(-\theta,n;\boldsymbol{\alpha})S^{\boldsymbol{\alpha}}_{kj}(\theta)=F_{\min}^{\mathcal{T}|jk+1}(2\pi i-\theta,n;\boldsymbol{\alpha}).
\end{align}
Similarly to the undeformed case, we have
\begin{align}
\label{eq:DFF-1}
F_{\min}^{\mathcal{T}|ii+k}(\theta,n;\boldsymbol{\alpha})&=F_{\min}^{\mathcal{T}|jj+k}(\theta,n;\boldsymbol{\alpha})\quad\forall\quad i,j,k  \\
\label{eq:DFF-2}
F_{\min}^{\mathcal{T}|1j}(\theta,n;\boldsymbol{\alpha})& =F_{\min}^{\mathcal{T}|11}(2\pi(j-1)i-\theta,n;\boldsymbol{\alpha})\quad\forall\quad j\neq1. 
\end{align}
Therefore, the form factor $F_{\min}^{T|11}(\theta,n;\boldsymbol{\alpha})$ is enough to determine all minimal form factors. 
\begin{align}
F_{\min}^{\mathcal{T}|11}(\theta,n;\boldsymbol{\alpha})=F_{\min}^{\mathcal{T}|11}(-\theta,n;\boldsymbol{\alpha})S_{\boldsymbol{\alpha}}(\theta)=F_{\min}^{\mathcal{T}|11}(-\theta+2\pi ni,n;\boldsymbol{\alpha}).
\end{align}
We assume the deformed minimal form factor takes a factorized form
\begin{align}
F_{\min}^{\mathcal{T}|11}(\theta,n;\boldsymbol{\alpha})=F_{\min}^{\mathcal{T}|11}(\theta,n)\varphi^{\mathcal{T}|11}(\theta,n;\boldsymbol{\alpha}),\quad \varphi^{\mathcal{T}|11}(\theta,n;0)=1\,.
\end{align}
Since the undeformed minimal form factor satisfies~\eqref{eq:undeformed F relation}, we have
\begin{align}
\varphi^{\mathcal{T}|11}(\theta,n;\boldsymbol{\alpha})=\varphi^{\mathcal{T}|11}(-\theta,n;\boldsymbol{\alpha})\Phi_{\boldsymbol{\alpha}}(\theta)=\varphi^{\mathcal{T}|11}(-\theta+2\pi n i,n;\boldsymbol{\alpha})\,.
\end{align}
The general solution is
\begin{align}
\varphi^{\mathcal{T}|11}(\theta,n;\boldsymbol{\alpha},\boldsymbol{\beta})=\exp\left(-\frac{i\pi-\theta/n}{2\pi}\sum_{s\in\mathcal{S}}\alpha_s\sinh s\theta\right)C(\theta,n;\boldsymbol{\beta})\,,
\end{align}
where the extra factor $C(\theta,n;\boldsymbol{\beta})$ is a new kind of ambiguity of the form factors, similar to the CDD factor for the S-matrix. More explicitly, it is given by
\begin{align}
\label{eq:FFCDD}
C(\theta,n;\boldsymbol{\beta})=\exp\left[\sum\limits_{s\in\mathcal{S'}}\beta_s\cosh \left(\frac{s\theta}{n}\right)\right].
\end{align} 
The generic CDD factor deformation has infinitely many parameters. In what follows, we focus on the $T\bar{T}$-deformation, which corresponds to taking $\alpha_s=\alpha\,\delta_{s,1}$. Even for this simple case, in principle we can have an infinite number of form factor CDD factors. Here we consider the effect of taking into account only one of them, corresponding to taking $\beta_s=\beta\,\delta_{1,s}$ in \eqref{eq:FFCDD}. In what follows, we shall refer to the solutions without such form factor CDD factors the \emph{simplest form factor} in order to distinguish it from the minimal form factor. We find this factor plays an interesting role for the positive deformation parameter $\alpha$ in $T\bar{T}$ deformation, which will be discussed in the next subsection. For the $T\bar{T}$ deformation, according to the relation~\eqref{eq:DFF-1} and~\eqref{eq:DFF-2},
\begin{align}
\varphi^{\mathcal{T}|jj}(\theta,n;\alpha,\beta)=&\varphi^{\mathcal{T}|11}(\theta,n;\alpha,\beta)=\exp \left[-\frac{\alpha(i \pi  n-\theta)\sinh\theta }{2 \pi  n}+\beta  \cosh \left(\frac{\theta}{n}\right)\right],\\
\varphi^{\mathcal{T}|1j}(\theta,n;\alpha,\beta)=&\exp \left[\frac{i \alpha  (n-j+1) \sinh\theta}{n}-\beta  \cosh \left(\frac{\theta}{n}\right)+\beta  \cosh \left(\frac{\theta-2 \pi i (j-1)}{n}\right)\right]\nonumber\\
&\times\varphi^{\mathcal{T}|11}(\theta,n;\alpha,\beta),\quad \forall\quad j\neq 1.
\end{align}
These relations allow us to determine all the deformed minimal form factors.
\paragraph{Deformed two-particle form factor} The full deformed 2-particle form factor can be obtained by taking into account the pole structures. Since the pole structure is given by the kinematic pole axioms, it is not modified. We expect poles at $\theta=i\pi$ and $\theta=i\pi(2n-1)$. This leads to the following full two-particle form factor
\begin{align}
F_2^{\mathcal{T}|jk}(\theta,n;\boldsymbol{\alpha},\boldsymbol{\beta})=&\frac{\langle \mathcal{T}\rangle\sin\left(\frac{\pi}{n}\right)}{2n\sinh\left(\frac{i\pi(2(j-k)-1)+\theta}{2n}\right)\sinh\left(\frac{i\pi(2(k-j)-1)-\theta}{2n}\right)}\frac{F_{\min}^{\mathcal{T}|jk}(\theta,n;\boldsymbol{\alpha})}{F_{\min}^{\mathcal{T}|jk}(i\pi,n;\boldsymbol{\alpha})}\\
=&F_2^{\mathcal{T}|jk}(\theta,n)\frac{\varphi^{\mathcal{T}|jk}(\theta,n;\boldsymbol{\alpha},\boldsymbol{\beta})}{\varphi^{\mathcal{T}|jk}(i\pi,n;\boldsymbol{\alpha},\boldsymbol{\beta})}.
\end{align}
Again for the $T\bar{T}$ deformation, the deformed two-particle form factor for branch-point twist fields also take a factorized form
\begin{align}
F_2^{\mathcal{T}|jk}(\theta,n;\alpha,\beta)=F_2^{\mathcal{T}|jk}(\theta,n)\frac{\varphi^{\mathcal{T}|jk}(\theta,n;\alpha,\beta)}{\varphi^{\mathcal{T}|jk}(i\pi,n;\alpha,\beta)}\,,
\end{align}
where 
\begin{align}
\varphi^{\mathcal{T}|jj}(i\pi,n;\alpha,\beta)&=\varphi^{\mathcal{T}|11}(i\pi,n;\alpha,\beta)=e^{\beta  \cos \left(\frac{\pi }{n}\right)},\\
\varphi^{\mathcal{T}|1j}(i\pi,n;\alpha,\beta)&=e^{\beta  \cos \left(\frac{(3-2 j)\pi}{n}\right)}.
\end{align}
\section{UV conformal dimension in deformed Ising model}
\label{sec:UVdim}
For the undeformed theory, in the UV limit, the conformal dimension of a primary field is related to a special correlation function~\cite{Delfino:1996nf}, which is the so-called $\Delta$-sum rule
\begin{align}
\label{eq:sumRule}
\Delta^{\mathcal{T}}=\Delta^{\tilde{\mathcal{T}}}=-\frac{1}{2\langle\mathcal{T}\rangle}\int_0^\infty r\langle\Theta(r)\tilde{\mathcal{T}}(0)\rangle dr,
\end{align}
where $\Theta$ is the trace of the stress-energy tensor. The scaling dimension is related to the conformal dimension by $d_n=2\Delta^{\mathcal{T}}$. The $\Delta$-sum rule has been studied in \cite{Castro-Alvaredo:2023rtl,Castro-Alvaredo:2023wmw,Castro-Alvaredo:2023jbg} and is shown to hold for local operators in the deformed theory. In this case, the $\Delta$-sum rule gives the deformed scaling dimension of $T\bar{T}$-deformed CFT. We make the assumption that it also holds for the branch-point twists fields and use it to obtain the deformed scaling dimension.
\par
The two-point function \eqref{eq:sumRule} can be calculated by the form factor expansion. We will approximate the expansion by the two-particle contribution
\begin{align}
\Delta^{\mathcal{T}}\approx-\frac{1}{2\langle\mathcal{T}\rangle}\sum_{i,j=1}^{n}\int_{-\infty}^{\infty}\int_{-\infty}^{\infty}\frac{d\theta_1d\theta_2F_2^{\Theta|ij}(\theta_{12})F_2^{\mathcal{T}|ij}(\theta_{12},n)^*}{2(2\pi)^2m^2\left(\cosh\theta_1+\cosh\theta_2\right)^2}\,,
\end{align}
where we have integrated out $r$. Since we consider the $n$ non-interacting copies, the form factor for $\Theta$ should be
\begin{align}
F_2^{\Theta|ii}(\theta)=F_2^{\Theta|11}(\theta),\quad F_2^{\Theta|ij}(\theta)=0\quad\text{for}\quad i\ne j.
\end{align}
Changing variables $\theta=\theta_1-\theta_2,\theta'=\theta_1+\theta_2$ and integrating out $\theta'$, one obtains
\begin{align}
\Delta^{\mathcal{T}}=-\frac{n}{32\pi^2m^2\langle\mathcal{T}\rangle}\int_{-\infty}^\infty d\theta\frac{F_2^{\Theta|11}(\theta)F_2^{\mathcal{T}|11}(\theta,n)^*}{\cosh^2(\theta/2)}.
\end{align}
To be more concrete, in what follows we will compute the deformed scaling dimension of the branch-point twist field for the $T\bar{T}$-deformed Ising model.
\par
For the Ising model, the two-particle form factors are given by
\begin{align}
F_2^{\Theta|11}(\theta)=&-2\pi im^2\sinh\left(\frac{\theta}{2}\right),\\
F_2^{\mathcal{T}|11}(\theta,n)=&\frac{-i\langle\mathcal{T}\rangle\cos\left(\frac{\pi}{2n}\right)}{n\sinh\left(\frac{i\pi+\theta}{2n}\right)\sinh\left(\frac{i\pi-\theta}{2n}\right)}\sinh\left(\frac{\theta}{2n}\right)\,.
\end{align}
After turning on the $T\bar{T}$-deformation, the deformed form factors become\footnote{{Note that the stress tensor form factor we used is different from the result in~\cite{Castro-Alvaredo:2023wmw,Castro-Alvaredo:2023jbg}, which reads
\begin{align}
\label{eq:stresstensor_FF_2}
F_2^{\Theta}(\theta;\boldsymbol\alpha)=-2\pi im^2\left|\frac{\sin\left(\sum_{s\in S}\frac{\alpha_s}{2}\sinh(s\theta)\right)}{\sum_ {s\in S}\frac{\alpha_s}{2}\sinh(s\theta)}\right|\sinh\frac{\theta}{2}\varphi(\theta;\boldsymbol\alpha).
\end{align}
However, adding an additional prefactor will not change much the result.}}
\begin{align}
F_2^{\Theta|11}(\theta;\alpha,\beta)=&F_2^{\Theta|11}(\theta)\frac{\varphi(\theta;\alpha,\beta)}{\varphi(i\pi;\alpha,\beta)},\\
F_2^{\mathcal{T}|11}(\theta,n;\alpha,\beta)=&F_2^{\mathcal{T}|11}(\theta,n)\frac{\varphi^{\mathcal{T}|11}(\theta,n;\alpha,\beta)}{\varphi^{\mathcal{T}|11}(i\pi,n;\alpha,\beta)},
\end{align}
where
\begin{align}
\varphi(\theta;\alpha,\beta)=&\exp\left[-\frac{\alpha(i\pi-\theta)}{2\pi}\sinh\theta+\beta\cosh\theta\right],\\
\varphi^{\mathcal{T}|11}(\theta,n;\alpha,\beta)=&\exp\left[-\frac{\alpha(in\pi-\theta)}{2n\pi}\sinh\theta+\beta\cosh \left(\frac{\theta}{n}\right)\right],\\
\varphi(i\pi;\alpha,\beta)=&e^{-\beta},\\
\varphi^{\mathcal{T}|11}(i\pi,n;\alpha,\beta)=&e^{\beta\cos\left(\frac{\pi}{n}\right)}.
\end{align}
\subsection{Simplest deformed form factor}
\begin{figure}%
\label{Figure:3.1}
\includegraphics[scale=0.7]{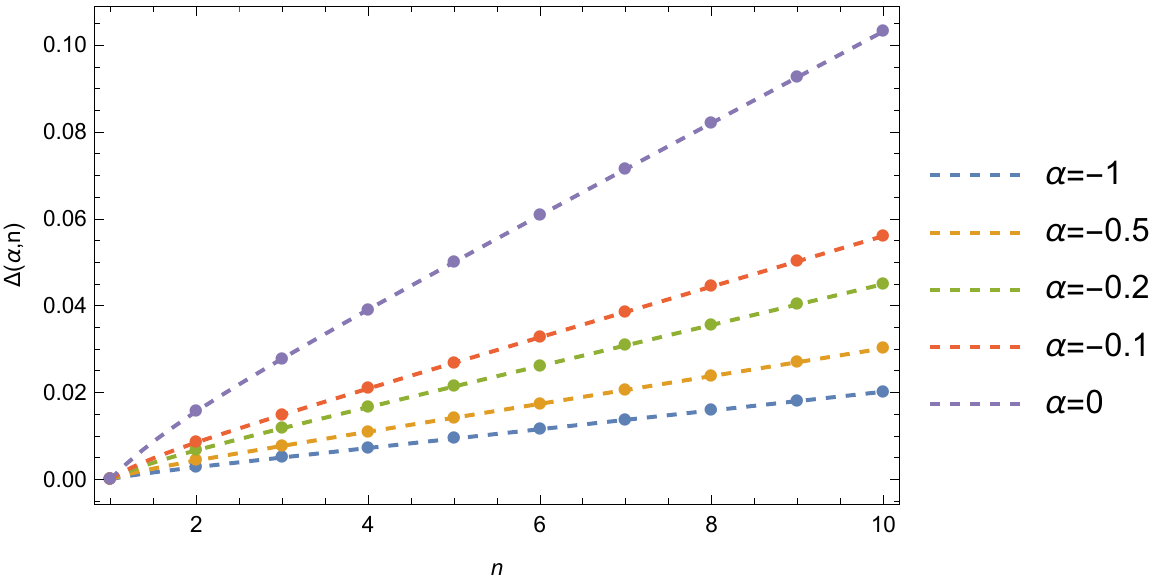}
\includegraphics[scale=0.7]{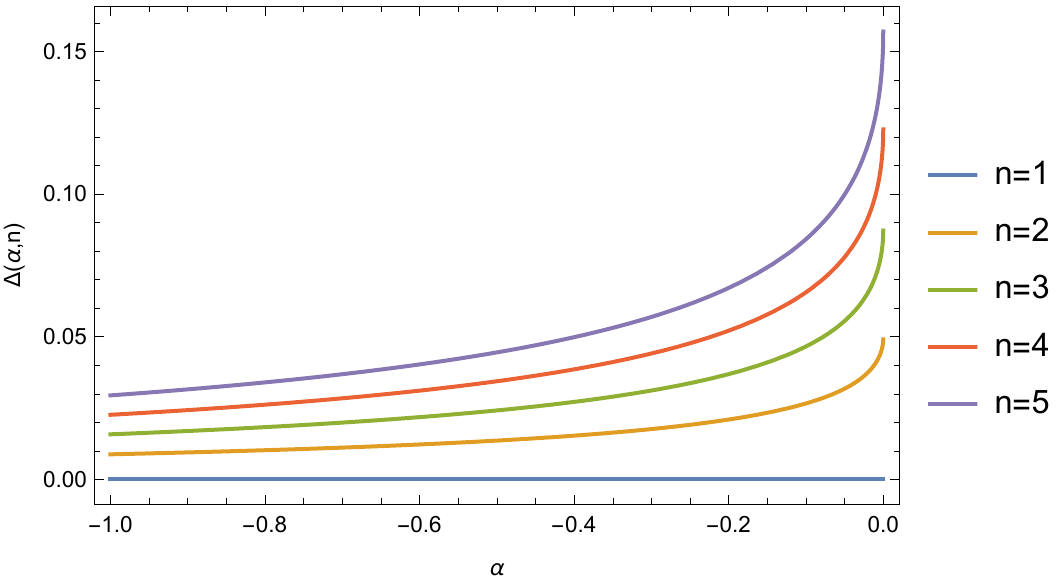}
\caption{Plot $\Delta(n;\alpha)$ as a function of $n$ and $\alpha$. The left figure shows that the conformal dimension becomes a linear function of $n$. For $n=1$, the conformal dimension  becomes vanish for negative $\alpha$. The right figure shows that the conformal dimension is always vanishing for $n=1$.}
\end{figure}

In this subsection, we shall first consider the case $\beta=0$, which corresponds to the simplest form factor. The deformed conformal dimension is given by
\begin{align}
\Delta(n;\alpha)&=-\frac{n}{32\pi^2m^2\langle\mathcal{T}\rangle_{\boldsymbol{\alpha}}}\int_{-\infty}^\infty d\theta\frac{F_2^{\Theta|11}(\theta)F_2^{\mathcal{T}|11}(\theta,n)^*}{\cosh^2(\theta/2)}\exp\left[\frac{\alpha\theta(n+1)}{2n\pi}\sinh\theta\right]\nonumber\\
&=-\frac{1}{8\pi}\int_{0}^\infty d\theta\frac{\cos\left(\frac{\pi}{2n}\right)\sinh\left(\frac{\theta}{2n}\right)\sinh\left(\frac{\theta}{2}\right)}{\sinh\left(\frac{i\pi+\theta}{2n}\right)\sinh\left(\frac{i\pi-\theta}{2n}\right)\cosh^2\left(\frac{\theta}{2}\right)}\exp\left[\frac{\alpha\theta(n+1)}{2n\pi}\sinh\theta\right].
\end{align}
Unfortunately, we are not able to work out the integral in a closed form, because of the exponential factor which comes from the $T\bar{T}$-deformation. We also find that the integral is convergent only for negative $\alpha$. We plot the figures of the conformal dimension as a function of $n$ and $\alpha$ in Figure~\ref{Figure:3.1}. 
From the figure, we find the conformal dimension becomes different for different deformation parameter $\alpha$, while they all approach to $0$ for $n=1$. This is one of the important features when we calculate the entanglement entropy. 
\par
In addition, we find the conformal dimension becomes a linear function of $n$ for large $n$. This feature was also found in~\cite{Castro-Alvaredo:2023jbg}. However, we need to study the behaviour of the function near $n=1$ since we have to take $n\to 1$ limit when calculating the entanglement entropy. Near $n=1$, the conformal dimension behaves like
\begin{align}
\Delta(\alpha,n)=A(\alpha)(n-1)+O\left((n-1)^2\right),
\end{align}
where
\begin{align}
\label{cd0}
A(\alpha)&=\int_{0}^{\infty}d\theta\sinh ^6\left(\frac{\theta }{2}\right) \text{csch}^4\theta \exp\left(\frac{\alpha  \theta  \sinh\theta}{\pi }\right).
\end{align}
For vanishing $\alpha$, we have
\begin{align}
A(0)=\frac{1}{24},
\end{align}
which corresponds to the undeformed case. Another important quantity is the derivative of the scaling dimension with respect to $n$. We have
\begin{align}
\label{con_d_0}
\Delta(n;\alpha)\Big|_{n=1}=0,\quad\frac{\partial\Delta(n;\alpha)}{\partial n}\Big|_{n=1}=A(\alpha).
\end{align}
\subsection{Taking into account form factor CDD factor }
\label{sec:EE}
In this subsection, we consider the case $\beta\neq 0$, which means we take the form factor CDD factor into account. The deformed scaling dimension is given by
\begin{align}
\label{CD}
\Delta(n;\alpha,\beta)
=&-\frac{1}{8\pi}\int_{0}^\infty d\theta\frac{\cos\left(\frac{\pi}{2n}\right)\sinh\left(\frac{\theta}{2n}\right)\sinh\left(\frac{\theta}{2}\right)}{\sinh\left(\frac{i\pi+\theta}{2n}\right)\sinh\left(\frac{i\pi-\theta}{2n}\right)\cosh^2\left(\frac{\theta}{2}\right)}\nonumber\\
&\times\exp\left[\frac{\alpha\theta(n+1)}{2n\pi}\sinh\theta+\beta\left(1-\cos\left(\frac{\pi}{n}\right)+\cosh\theta+\cosh\left(\frac{\theta}{n}\right)\right)\right].
\end{align}
In this case, there is another exponential term in the integrand. The convergence of this integral is determined by the competition between these two terms in the exponent. For the simplest form factor, the integral is convergent just for negative $\alpha$. However, the $T\bar{T}$ deformation for positive $\alpha$ is also interesting which relates to the holography~\cite{McGough:2016lol}. We find the appearance of the form factor CDD factor would lead to the convergence of this integral with positive $\alpha$. We then discuss the convergence of the integral in the following:
\begin{itemize}
\item $\alpha<0$
\end{itemize}
\par
\begin{figure}%
\label{Figure:3.2}
\includegraphics[scale=0.7]{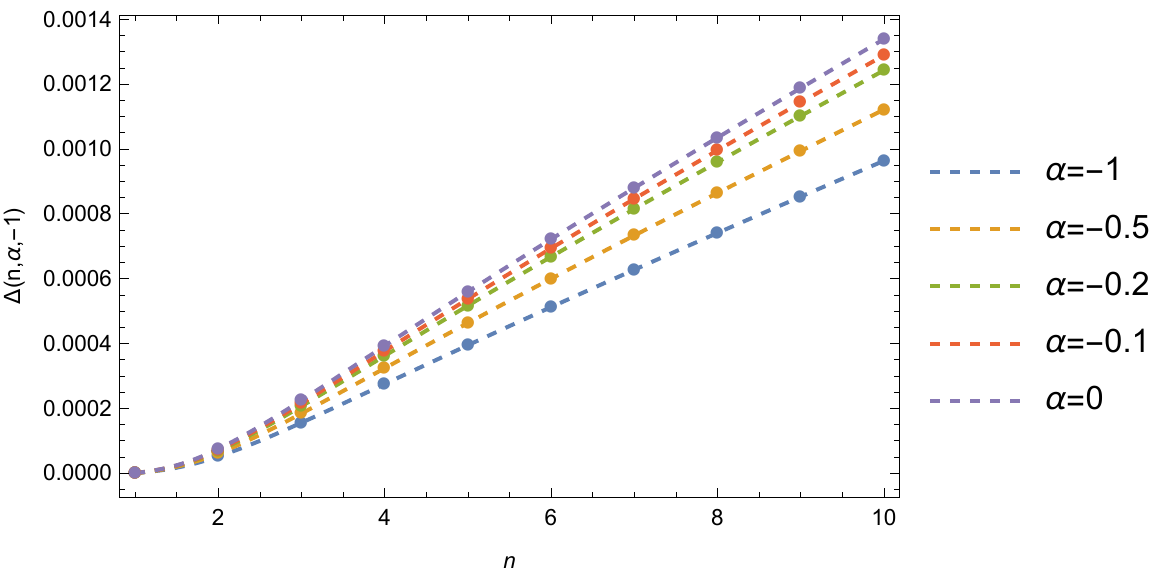}
\includegraphics[scale=0.7]{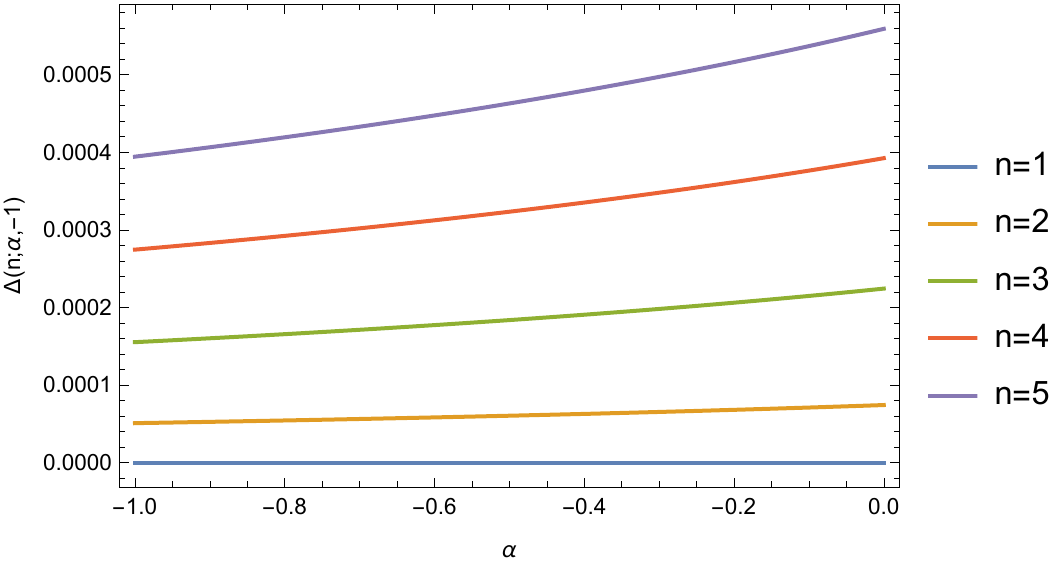}
\caption{Plot of $\Delta(n;\alpha,\beta)$ as a function of $n$ and $\alpha$ with $\beta=-1$. The left figure shows that the conformal dimension becomes a linear function of $n$ for large $n$. The right figure shows that the conformal dimension is always vanishing for $n=1$.}
\end{figure}

The integral is always convergent in this case, because we have $\theta\sinh\theta\gg \cosh\theta$ for large $\theta$ and the $\alpha$-dependent term is the dominating term in the exponential factor. We study the $\alpha$ dependence of the scaling dimension for a fixed $\beta$. We plot the scaling dimension with $\beta=-1$ in Figure~\ref{Figure:3.2}. Similar behaviours can also be found for positive $\beta$, but the value of conformal dimension is very large. For the small $n$, we have
\begin{align}
\Delta(n;\alpha,\beta)=A_{\beta}(\alpha)(n-1)+O\left((n-1)^2\right),
\end{align}
where
\begin{align}
\label{cd1}
A_{\beta}(\alpha)=\int_{0}^{\infty}d\theta\sinh ^6\left(\frac{\theta }{2}\right) \text{csch}^4\theta \exp\left[\frac{\alpha \theta \sinh\theta}{\pi }+2 \beta  (\cosh \theta+1)\right].
\end{align}
Moreover, we have
\begin{align}
\label{con_d_1}
\Delta(n;\alpha,\beta)\Big|_{n=1}=0,\quad\frac{\partial\Delta(n;\alpha,\beta)}{\partial n}\Big|_{n=1}=A_{\beta}(\alpha).
\end{align}
\begin{itemize}
\item $\alpha>0$
\end{itemize}
The integral is always divergent for the simplest form factor when $\alpha>0$. With the presence of the form factor CDD factor contribution, the integral is convergent provided
\begin{align}
\label{convergence_condition}
\frac{\alpha\theta(n+1)}{2n\pi}\sinh\theta+\beta\left(1-\cos\left(\frac{\pi}{n}\right)+\cosh\theta+\cosh\left(\frac{\theta}{n}\right)\right)<0.
\end{align}
The above inequality holds only if $\beta<0$. As the rapidity increases, the first term on the left side will eventually become dominant and it would break the inequality. Therefore the inequality~\eqref{convergence_condition} gives an upper bound of the rapidity for the integral to be convergent. Basically, we have a UV cut-off of the rapidity to ensure the convergence of the integral~\eqref{CD} and get a well-defined conformal dimension.

For large $\theta\gg 1$, the inequality above can be approximated by
\begin{align}
\frac{\alpha(n+1)\theta e^{\theta}}{4n\pi}+\frac{\beta}{2} \left(e^{\frac{\theta}{n}}+e^{\theta}\right)<0,
\end{align}
so we have the rapidity cut-off
\begin{align}
\Lambda_n=&\frac{n}{n-1}W_0\left(-\frac{2 \pi  \beta  (n-1) }{\alpha  (n+1)}e^{\frac{2 \pi  \beta  (n-1)}{\alpha  (n+1)}}\right)-\frac{2 \pi  \beta  n}{\alpha(1+n)}.
\end{align}
The rapidity cut-off is given by the Lambert $W$-function, which  shows up in various contexts related to $T\bar{T}$ deformation, such as hard-rod deformation as well as $T\bar{T}$-deformation~\cite{Cardy:2020olv,Jiang:2020nnb} of Bose gas. In~\cite{Castro-Alvaredo:2023rtl,Castro-Alvaredo:2023wmw}, the rapidity cut-off was also introduced when they study the correlation functions in $T\bar{T}$-deformed Ising model. After introducing the rapidity cut-off, we obtain the convergent conformal dimension
\begin{align}
\Delta^{\text{cut}}(n;\alpha,\beta)
=&-\frac{1}{8\pi}\int_{0}^{\Lambda_n} d\theta\frac{\cos\left(\frac{\pi}{2n}\right)\sinh\left(\frac{\theta}{2n}\right)\sinh\left(\frac{\theta}{2}\right)}{\sinh\left(\frac{i\pi+\theta}{2n}\right)\sinh\left(\frac{i\pi-\theta}{2n}\right)\cosh^2\left(\frac{\theta}{2}\right)}\nonumber\\
&\times\exp\left[\frac{\alpha\theta(n+1)}{2n\pi}\sinh\theta+\beta\left(1-\cos\left(\frac{\pi}{n}\right)+\cosh\theta+\cosh\left(\frac{\theta}{n}\right)\right)\right].
\end{align}
The scaling dimension exhibit similar behaviour to the positive $\alpha$ case, which is plotted in Figure 3.3.
\begin{figure}
\label{Figure:3.3}
\includegraphics[scale=0.7]{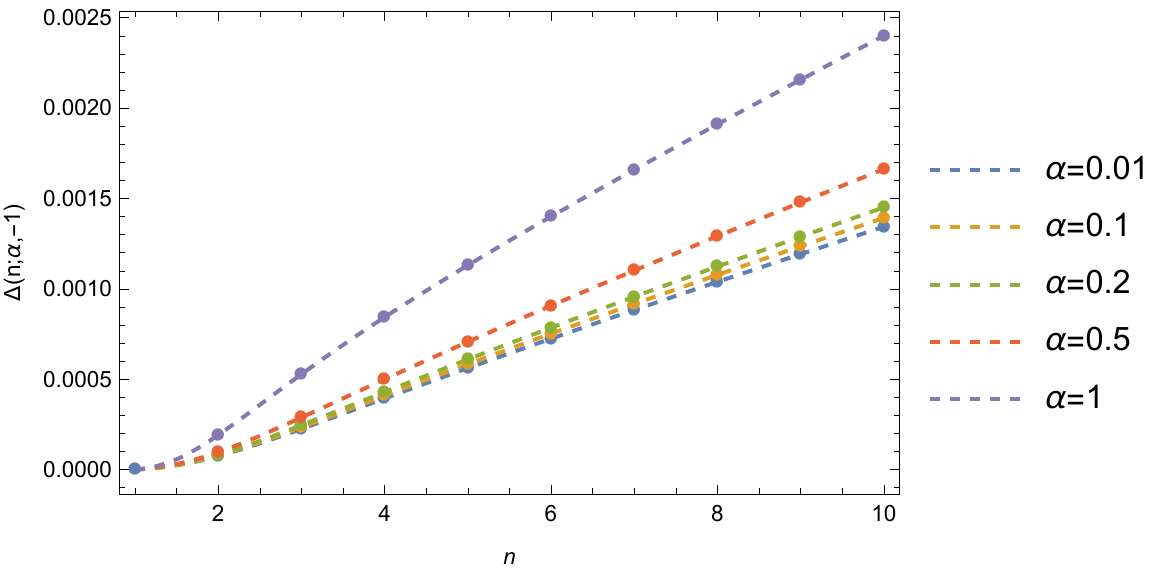}
\includegraphics[scale=0.7]{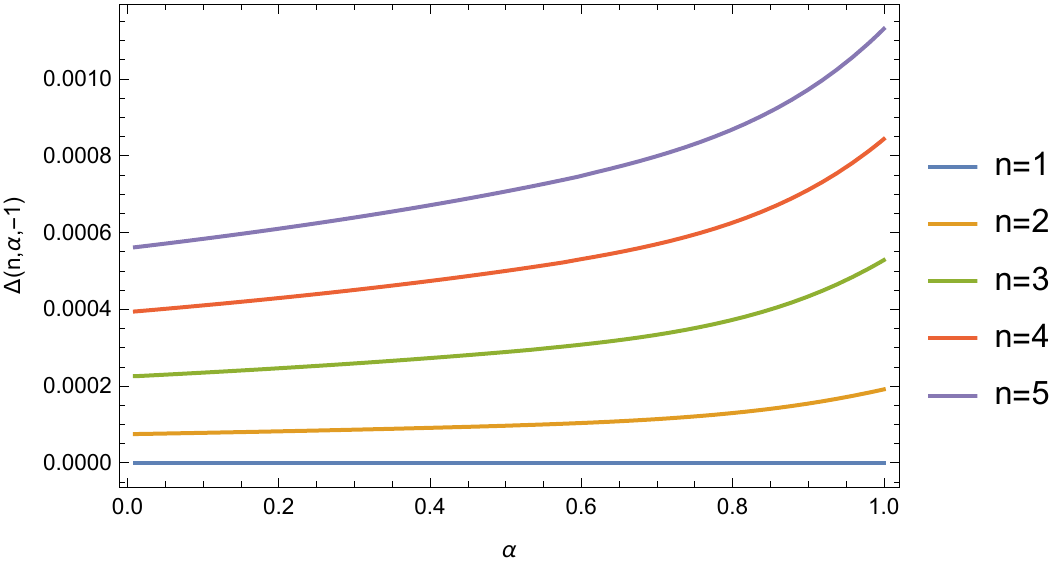}
\caption{Plot the $\Delta^{\text{cut}}(n;\alpha,\beta)$ as a function of $n$ and $\alpha$ with $\beta=-1$. The left figure shows that the conformal dimension becomes a linear function of $n$ for large $n$. The right figure shows that the conformal dimension is always vanishing for $n=1$.}
\end{figure}
\par
In the $n\to 1$ limit, the rapidity cut-off becomes 
\begin{align}
\Lambda_1=-\frac{2 \pi  \beta }{\alpha }.
\end{align}
The scaling dimension in this limit can be expanded as
\begin{align}
\Delta^{\text{cut}}(n;\alpha,\beta)=A^{\text{cut}}_{\beta}(\alpha)(n-1)+O\left((n-1)^2\right),
\end{align}
where
\begin{align}
\label{cd2}
A^{\text{cut}}_{\beta}(\alpha)=\int_{0}^{-\frac{2\pi\beta}{\alpha}}d\theta \sinh ^6\left(\frac{\theta}{2}\right) \text{csch}^4\theta\exp\left[\frac{\alpha  \theta\sinh\theta}{\pi }+2 \beta  (\cosh\theta+1)\right]\,.
\end{align}
We have
\begin{align}
&\Delta^{\text{cut}}(n;\alpha,\beta)\Big|_{n=1}=0,\quad \frac{\Delta^{\text{cut}}(n;\alpha,\beta)}{\partial {n}}\Big|_{n=1}=A_{\beta}^{\text{cut}}(\alpha).
\end{align}
\section{$T\bar{T}$-deformed entanglement entropy}
\label{sec:EE}
We now can compute the deformed entanglement entropy. Without loss of generality, we place the two branch-point twist fields at $(t,x)=(0,0)$ and $(t,x)=(0,r)$. The two-point function of the branch-point twist fields, up to two-particle approximation is given by
\begin{align}
\langle\mathcal{T}(r)\tilde{\mathcal{T}}(0)\rangle {\approx}&{\langle\mathcal{T}\rangle^{2}+\sum_{i,j=1}^{n}\int_{-\infty}^{\infty}\int_{-\infty}^{\infty}\frac{d\theta_{1}d\theta_{2}}{2!(2\pi)^{2}}\left|F_{2}^{\mathcal{T}|ij}(\theta_{12},n)\right|^{2}\left|\frac{\varphi^{\mathcal{T}|ij}(\theta_{12},n;\alpha,\beta)}{\varphi^{\mathcal{T}|ij}(i\pi,n;\alpha,\beta)}\right|^2e^{-r m(\cosh\theta_{1}+\cosh\theta_{2})}}
\end{align}
Introducing the new variables $\theta=\theta_1-\theta_2,\theta'=\theta_1+\theta_2$ and integrating out $\theta'$,
one obtains
\begin{align}
\langle\mathcal{T}(r)\tilde{\mathcal{T}}(0)\rangle 
=&\langle\mathcal{T}\rangle^{2}\left(1+\frac{n}{4\pi^2}\int_{-\infty}^{\infty}d\theta f(\theta,n;\alpha,\beta)K_0\left(2mr\cosh\frac{\theta}{2}\right)\right),
\end{align}
where 
\begin{align}
K_0(z)=&\int_0^\infty e^{-z\cosh t}dt
\end{align}
is the Bessel function and we have used the relation
\begin{align}
\label{definition_f}
&f(\theta,n;\alpha,\beta)=\sum_{j=1}^n\left|F_2^{\mathcal{T}|1j}(\theta,n)\right|^2\left|\frac{\varphi^{\mathcal{T}|1j}(\theta,n;\alpha,\beta)}{\varphi^{\mathcal{T}|1j}(i\pi,n;\alpha,\beta)}\right|^2\nonumber\\
=&\sum_{j=1}^n\left|F_2^{\mathcal{T}|1j}(\theta,n)\right|^2\exp \left(-2 \beta  \cos \left(\frac{\pi  (3-2 j)}{n}\right)+2 \beta  \cos \left(\frac{2 \pi  (j-1)}{n}\right) \cosh\frac{\theta}{n}+\frac{\alpha\theta\sinh\theta}{\pi  n}\right).
\end{align}
The two-point function is a function of $n$ and $mr$. To obtain the entanglement entropy we have to perform analytic continuation of~\eqref{definition_f} and then evaluate the derivative at $n=1$.  The following results for the undeformed case~\cite{Cardy:2007mb} are useful 
\begin{align}
\lim_{n\to 1}f(\theta,n)=0,\quad \lim_{n\to 1}\frac{\partial}{\partial n}f(\theta,n)=&\frac{\pi^2}{2}\delta(\theta)\,,
\end{align}
where
\begin{align}
f(\theta,n)=\sum_{j=1}^n\left|F_2^{\mathcal{T}|1j}(\theta,n)\right|^2.
\end{align}
We find the deformation contributes an exponential factor in~\eqref{definition_f}, which depends on the summation index $j$. Fortunately, the exponential factor becomes $j$-independent when taking $n\to 1$ limit. We then obtain the following results
\begin{align}
&\lim_{n\to 1}f(\theta,n;\alpha,\beta)=0,\\
&\lim_{n\to 1}\frac{\partial}{\partial n}f(\theta,n;\alpha,\beta)=\frac{\pi^2}{2}\delta(\theta)\exp\left(\frac{\alpha\theta\sinh\theta}{\pi }+2 \beta  (\cosh \theta+1)\right).
\end{align}
Finally, the entanglement entropy can be calculated by
\begin{align}
S(mr,\alpha,\beta)=&-\lim_{n\to 1}\frac{\partial}{\partial n}\left(Z_n\epsilon^{4\Delta(n;\alpha,\beta)}\langle\mathcal{T}(r)\tilde{\mathcal{T}}(0)\rangle\right).
\end{align}
\par
For the simplest deformed form factor, the entanglement entropy reads
\begin{align}
S(mr,\alpha)=-\frac{c(\alpha)}{3}\log(m \epsilon)+U(\alpha)-\frac{1}{8}K_0(2mr)+...
\end{align}
where
\begin{align}
U(\alpha)=-\frac{\partial}{\partial n}\left(m^{-4\Delta(n;\alpha)}\langle \mathcal{T}\rangle^2\right)\Big|_{n=1},\quad c(\alpha)=12A(\alpha).
\end{align}
and $A(\alpha)$ is defined in~\eqref{cd0}. This is the same result as~\cite{Castro-Alvaredo:2023jbg}~\footnote{We obtained a different result in the first version of this paper. The reason is a miscalculation of the deformed conformal dimension. We thank the authors of~\cite{Castro-Alvaredo:2023jbg} for pointing this out.}. The UV behaviour of the entanglement entropy takes the same form as the undeformed case, except for the change of central charge. Since the UV conformal dimension is well-defined for negative $\alpha$, the entanglement entropy formula holds only for $\alpha<0$ in this case. 
\par 
Including the form factor CDD factor, for $\alpha<0$, the entanglement entropy reads
\begin{align}
S(mr,\alpha,\beta)=-\frac{c(\alpha,\beta)}{3}\log(m \epsilon)+U(\alpha,\beta)-\frac{1}{8}e^{4\beta}K_0(2mr)+...
\end{align}
where
\begin{align}
U(\alpha,\beta)=-\frac{\partial}{\partial n}\left(m^{-4\Delta(n;\alpha,\beta)}\langle \mathcal{T}\rangle^2\right)\Big|_{n=1},\quad c(\alpha,\beta)=12A_{\beta}(\alpha).
\end{align}
and $A_{\beta}(\alpha)$ is defined in~\eqref{cd1},
In this case, the entanglement entropy becomes $\beta$ dependent for both UV and IR corrections. The result can reduce to the simplest deformed form factor result after taking $\beta\to 0$.
\par
For $\alpha>0$ and $\beta<0$, the entanglement entropy becomes
\begin{align}
S(mr,\alpha,\beta)=-\frac{c^{\text{cut}}(\alpha,\beta)}{3}\log(m \epsilon)+U^{\text{cut}}(\alpha,\beta)-\frac{1}{8}e^{4\beta}K_0(2mr)+...
\end{align}
where
\begin{align}
U^{\text{cut}}(\alpha,\beta)=-\frac{\partial}{\partial n}\left(m^{-4\Delta^{\text{cut}}(n;\alpha,\beta)}\langle \mathcal{T}\rangle^2\right)\Big|_{n=1},\quad c^{\text{cut}}(\alpha,\beta)=12A^{\text{cut}}_{\beta}(\alpha),
\end{align}
and $A^{\text{cut}}_{\beta}(\alpha)$ is defined in~\eqref{cd2}.
\par
We find the UV part of the deformed entanglement entropy behaves like undeformed one except for the deformation of central charge which now depends on both $\alpha$ and $\beta$. These results disagree with the perturbative calculation~\cite{Ashkenazi:2023fcn}, in which they found a log-square term in the first order correction. In~\cite{Castro-Alvaredo:2023jbg}, the authors discussed several possibilities to reproduce the such terms such as adjusting the vacuum expectation value of the twist operator. Here we give another possibility.
For the $\alpha>0$ and $\beta<0$ case, the rapidity cut-off $\Lambda_1$ is introduced. In fact, there is also a short distance cut-off $\epsilon$ in the entanglement entropy formula. A priori, these two cut-offs are independent. However, things become interesting if we take $\beta$ to be the same scale as $\epsilon$. The momentum is related to the rapidity by
\begin{align}
p=m\sinh\theta\sim \frac{m}{2}e^{\theta},\quad\text{for}\quad\theta\gg 1.
\end{align}
If the short distance cut-off $\epsilon$ is introduced, the momentum UV cut-off should be $p\sim 1/\epsilon$. Therefore, the rapidity cut-off can relate to the short distance cut-off by
\begin{align}
\label{cutoff_relation}
\frac{m}{2}e^{\Lambda_1}\sim\frac{1}{\epsilon},
\end{align}  
which leads to
\begin{align}
\beta\sim \frac{\alpha}{2\pi}\log\left(\frac{m\epsilon}{2}\right).
\end{align}
Therefore we can replace the parameter $\beta$ by the cut-off $\epsilon$. The UV part of the $T\bar{T}$-deformed entanglement entropy becomes 
\begin{align}
&S(mr,\alpha)\nonumber\\
=&-4\log(m\epsilon)\int_0^{\log\left(\frac{2}{m\epsilon}\right)}d\theta \sinh ^6\left(\frac{\theta}{2}\right) \text{csch}^4\theta\exp\left[\frac{\alpha \theta\sinh\theta}{\pi }+\frac{\alpha}{\pi}\log\left(\frac{m\epsilon}{2}\right) (\cosh\theta+1)\right]\nonumber\\
\sim&-\frac{1}{6}\log(m\epsilon)+\left(\frac{\pi ^2+3 (2-\log2)^2}{12 \pi }+\frac{(2-\log2) \log (m \epsilon )}{2 \pi }+\frac{\log^2 (m \epsilon )}{4\pi }\right)\log(m\epsilon)\alpha+O(\alpha^2)
\end{align}
To obtain the perturbation result, we first take the perturbation expansion of $\alpha$ then take $\epsilon\to 0$ limit. In fact, we are working with the double-scaling limit
\begin{align}
\epsilon\to 0,\quad \alpha\to 0,\quad \alpha\log\left(m\epsilon\right)=\text{fixed}.
\end{align} 
In this sense, the leading order reproduces the undeformed entanglement entropy and the first order correction has richer structure, consistent with perturbative results~\cite{Ashkenazi:2023fcn}.
\section{Conclusions and outlook}
\label{sec:conlusion}
In this work, we computed the deformed von Neumann entropy $T\bar{T}$-deformed IQFTs. For simplicity, we have focused on the diagonal scattering theory with a single particle and no bound states. We obtained the form factors of the branch-point twist fields by solving the form factor bootstrap axioms for the $T\bar{T}$-deformed theories. The solution is not unique and we investigated two types of solutions. The first type is the simplest solution, we find that the UV behavior of the deformed entanglement entropy is modified in a simple way, which is just a change of the central charge. The other case which we consider is multiplying the simplest solution by a CDD-like factor, which still solves the bootstrap axioms. In this case, we have two parameters. Depending on the interplay of the two parameters, the resulting two-particle contribution can be fully convergent, or convergent up to a certain cut-off rapidity. In the latter case, if we identify this cut-off with the UV cut-off that is intrinsic to the calculation of the entanglement entropy, the UV limit of the entanglement entropy can have different behaviors. Our result is non-perturbative in the deformation parameter and is the first non-perturbative result on deformed entanglement entropy for the $T\bar{T}$-deformed massive QFTs.\par

There are many future directions to pursue. To start with, our result so far is constrained in the two-particle sector. Although it is already sufficient to see the main features, it would be desirable to have higher particle form factors \cite{Castro-Alvaredo:2011hnn} and push the computations further. Second, more general theories such as the ones with multiple particle types, bound states and non-diagonal scatterings should be investigated. Finally, theories with integrable defects \cite{Delfino:1994nx,Delfino:1994nr,Jiang:2017qhn,Jiang:2017tyi} and boundaries \cite{Ghoshal:1993tm,Castro-Alvaredo:2008fni} are also interesting to study. $T\bar{T}$- and more general deformed theories for boundary IQFTs have been initiated in \cite{Jiang:2021jbg}.\par

There are other types of branch-point twist fields which can be used to compute other quantum information measures, such as the logarithmic negativity \cite{Blondeau-Fournier:2015yoa} and symmetry resolved entanglement entropy \cite{Capizzi:2022jpx,Capizzi:2022nel,Capizzi:2023ksc}. It would be valuable to study the behavior of these quantum information measures under the CDD factor deformations.\par

As we have already stressed in the introduction. It would be important to further justify the applicability of form factor bootstrap axioms for $T\bar{T}$-deformed IQFTs. For local IQFTs, these axioms can be derived from LSZ reduction formula and the maximal analyticity assumption \cite{Babujian:1998uw} while for $T\bar{T}$-deformed theories such a proof is lacking. It is also crucial to understand the asymptotic behaviors of the deformed two-point functions, which will help us to fix the correct form factor.

Finally it would be interesting to see if the results or at least some of the key ideas can be generalized to other integrable models such as quantum spin chains and 1D cold atom system. It is of great interest to study deformed correlation functions and quantum information measures for such models.

\section*{Acknowledgments}
We would like to thank the authors of~\cite{Castro-Alvaredo:2023jbg}, in which they point out a miscalculation in the first version of this paper. MH is supported by China Postdoctoral Science Foundation under Grant No.2023M740612.

\providecommand{\href}[2]{#2}\begingroup\raggedright\endgroup
\end{document}